# Enhancement of Coulomb blockade in epsilon near zero and hyperbolic metamaterials

Igor I. Smolyaninov [1,2)], Vera N. Smolyaninova [3)]

[1]Department of Electrical and Computer Engineering, University of Maryland, College Park, MD 20742, USA

[2]Saltenna LLC, 1751 Pinnacle Drive #600 McLean, VA 22102,USA

[3]Department of Physics Astronomy and Geosciences, Towson University,

8000 York Rd., Towson, MD 21252, USA

**Coulomb interaction of charges in a metamaterial may be expressed via its effective dielectric response function. Coulomb interaction appears considerably enhanced in artificially engineered epsilon near zero (ENZ) and hyperbolic metamaterials. Indeed, very recently it was observed that negative ENZ conditions enhance superconducting properties of a composite metamaterial via enhancement of Cooper pairing. Here we demonstrate that positive ENZ conditions enhance Coulomb repulsion of charges, leading to considerable enhancement of the Coulomb blockade effect in such metamaterials. Similar to high Tc cuprates, in hyperbolic metamaterials superconductivity and the Coulomb blockade-induced charge ordering is predicted to compete with each other.**

Within the scope of macroscopic electrodynamics the Coulomb interaction of charges in a metamaterial may be expressed via its effective dielectric response function $\varepsilon_{eff}(q,\omega)$:

$$V(\vec{q},\omega) = \frac{4\pi e^2}{q^2 \varepsilon_{eff}(\vec{q},\omega)} = \frac{V_C}{\varepsilon_{eff}(\vec{q},\omega)} \,, \quad (1)$$

where $V_C = 4\pi e^2/q^2$ is the Fourier-transformed Coulomb potential in vacuum [1]. Coulomb interaction of charges appears to be considerably enhanced in artificially engineered epsilon near zero (ENZ) [2] and hyperbolic [3] metamaterials. Indeed, very recently it was observed that negative ENZ conditions enhance superconducting properties of a composite metamaterial via enhancement of Cooper pairing. For example, tripling of the superconducting critical temperature has been observed in Al-$Al_2O_3$ ENZ core-shell and hyperbolic metamaterials [4,5]. On the other hand, it may be expected that positive ENZ conditions will enhance Coulomb repulsion of charges, leading to enhancement of the Coulomb blockade effect in metamaterials. An artificially engineered metamaterial in which $\varepsilon_{eff}(q,\omega)$ is close to zero (while remaining positive) will experience considerable enhancement of repulsive electron-electron interaction, leading to enhancement of the Coulomb blockade effect and therefore to considerable enhancement of such practically important effects as single electron tunnelling, spin blockade, valley blockade, etc. While standard Coulomb blockade theory [6-8] considers circuits with constant voltage bias, more recent work [9-11] has addressed nanostructures driven by alternating voltages inspired by nanoelectronic experiments in the GHz range and above [12,13]. Therefore, application of concepts developed in the field of electromagnetic metamaterials appears to be highly appropriate in the case of AC driven dynamical Coulomb blockade effect.

Currently developed Coulomb blockade-based electronic devices (see Fig.1a), such as single electron transistors (SETs) and spintronic devices based on spin blockade need to have either extremely small nanometer-scale dimensions (leading to extremely small self-capacitance $C$) or must be cooled to near zero temperatures, so that $k_BT < e^2/C$ condition would be fulfilled [14]. A positive ENZ metamaterial implemented as shown in Fig.1b may help overcome these limitations by making self-capacitance very small even in macroscopically large devices, so that AC driven SETs and spin blockade-based spintronic devices may be operated at room temperature. Such positive ENZ metamaterials may be fabricated using the same guiding principles as in fabrication of negative ENZ metamaterial superconductors. For example, carefully designed mixtures of positive and negative $\varepsilon$ materials may be implemented, as well as hyperbolic metamaterial designs [4,5]. In this paper we are going to consider several such designs and derive metamaterial structural parameters necessary to achieve the dynamic Coulomb blockade effect in the GHz and THz ranges. We also demonstrate that in hyperbolic metamaterials the positive and negative ENZ conditions appear to coexist depending on the spatial direction, leading to competition between the charge density waves and the superconducting behaviour in such metamaterials.

First, let us consider isotropic ENZ designs, which may be achieved using either random mixtures of positive and negative $\varepsilon$ constituents (Fig.2a) [15], or using core-shell nanoparticle geometries (Fig.1b) [4]. According to the Maxwell-Garnett approximation [16], random mixing of nanoparticles of a metal "matrix" with dielectric "inclusions" (described by the dielectric constants $\varepsilon_m$ and $\varepsilon_i$, respectively) results in the effective medium with a dielectric constant $\varepsilon_{eff}$, which may be obtained as

$$\left(\frac{\varepsilon_{eff} - \varepsilon_m}{\varepsilon_{eff} + 2\varepsilon_m}\right) = \delta_i \left(\frac{\varepsilon_i - \varepsilon_m}{\varepsilon_i + 2\varepsilon_m}\right), \qquad (2)$$

where $\delta_i$ is the volume fraction of the inclusions. This expression is valid as long as $\delta_i$ is small. The explicit expression for $\varepsilon_{eff}$ may be written as

$$\varepsilon_{eff} = \varepsilon_m \frac{(2\varepsilon_m + \varepsilon_i) - 2\delta_i(\varepsilon_m - \varepsilon_i)}{(2\varepsilon_m + \varepsilon_i) + \delta_i(\varepsilon_m - \varepsilon_i)} \qquad (3)$$

The ENZ condition ($\varepsilon_{eff} \approx 0$) are obtained around

$$\delta_i = \frac{2\varepsilon_m + \varepsilon_i}{2(\varepsilon_m - \varepsilon_i)}, \qquad (4)$$

which means that $\varepsilon_m$ and $\varepsilon_i$ must have opposite signs, and $\varepsilon_i \approx -2\varepsilon_m$ so that $\delta_i$ given by Eq.(4) will be small. Operating at $\delta_i$ above the critical value given by Eq.(4) ensures positive ENZ conditions.

While random nanoparticle mixture geometry clearly provides a pathway to fabricate positive ENZ metamaterials, it is not ideal. It is clear that simple mixing of metal and dielectric nanoparticles leaves room for substantial spatial variations of $\delta_i$ throughout a macroscopic metamaterial sample. This issue may be resolved by implementing a core-shell metamaterial geometry shown schematically in Fig.2b. The design of an individual core-shell nanoparticle is based on the fact that scattering of electromagnetic field by a subwavelength object is dominated by its electric dipolar contribution, which is defined by the integral sum of its volume polarization [17]. A material with $\varepsilon > 1$ has a positive electric polarizability, while a material with $\varepsilon < 1$ has a negative electric polarizability (since the local electric polarization vector $P=(\varepsilon-1)E/4\pi$ is opposite to $E$). As a result, the presence of a metal shell cancels the scattering produced by a dielectric core. Similar consideration for the positive ENZ case leads to the following condition for the core-shell geometry:

$$r_1^3 \varepsilon_i \approx -(r_2^3 - r_1^3)\varepsilon_m, \qquad (5)$$

where $r_1$ and $r_2$ are the radii of the dielectric core and the metal shell, respectively. Eq.(5) corresponds to average dielectric permittivity of the core-shell nanoparticle being approximately equal to zero. Working on the positive side of this equality will ensure positive ENZ character of each core-shell nanoparticle. A dense assembly of such core-shell nanoparticles will form a medium, which will have small positive dielectric permittivity. Once again, $\varepsilon_m$ and $\varepsilon_i$ must have opposite signs and the same order of magnitude for the core-shell metamaterial design to work as a positive ENZ metamaterial.

Dielectric permittivity $\varepsilon_m = \varepsilon_m(0,\omega)$ of the metal component is typically given by the Drude model in the far infrared and THz ranges as

$$\varepsilon_m = \varepsilon_{m\infty} - \frac{\omega_p^2}{\omega(\omega+i\gamma)} \approx -\frac{\omega_p^2}{\omega(\omega+i\gamma)}, \quad (6)$$

where $\varepsilon_{m\infty}$ is the dielectric permittivity of metal above the plasma edge, $\omega_p$ is its plasma frequency, and $\gamma$ is the inverse free propagation time. Since $\varepsilon_m$ is large and negative, the dielectric permittivity $\varepsilon_i$ of dielectric inclusions (or dielectric cores) must be positive and very large. Ferroelectric materials, such as $BaTiO_3$, which have large positive $\varepsilon_i$ in the THz range appear to be a very good choice of such a dielectric. The high frequency behavior of $\varepsilon_i$ may be assumed to follow the Debye model [18]:

$$\mathrm{Re}\,\varepsilon_i = \frac{\varepsilon_i(0)}{1+\omega^2\tau^2} \approx \frac{\varepsilon_i(0)}{\omega^2\tau^2}, \quad (7)$$

resulting in a broadband positive ENZ behaviour of the metal-dielectric metamaterial due to the similar $\sim\omega^{-2}$ functional behaviour of $\varepsilon_i$ and $\varepsilon_m$ in the THz range (see Eqs. (6) and (7)).

We have fabricated several nanoparticle-based metal-dielectric metamaterials of varying composition and measured their dielectric properties as illustrated in Figs. 2a and 3 and described in detail in [15]. These metamaterial samples were prepared using commercially available tin and barium titanate nanoparticles obtained from the US Research Nanomaterials, Inc. The nominal diameter of the $BaTiO_3$ nanoparticles was 50 nm, while tin nanoparticle size was specified as 60-80 nm. Our choice of materials was based on results of numerical calculations of the real part of the dielectric constant of the Sn/BaTiO$_3$ mixture shown in Fig.3a. These calculations were based on the measured dielectric properties of Sn [19] and BaTiO$_3$ [20] extrapolated by Eqs. (6) and (7), respectively, and on the Maxwell-Garnett expression, Eq. (3), for the dielectric permittivity of the mixture. These calculations indicated that it was possible to achieve broadband positive ENZ conditions in the 30-50% range of the volume fraction of BaTiO$_3$. Measurements performed on the 40% mixture shown in Fig.3b demonstrate validity of this approach. A positive ENZ material may be deposited on top of the island between the source and the drain of the Coulomb blockade device as shown in Fig.1b. When the island is surrounded by the positive ENZ metamaterial, its self-capacitance is greatly reduced leading to much more relaxed nanofabrication requirements, or the ability to operate such device at much higher temperatures. We should also note that low loss positive ENZ behavior is observed in low carrier concentration semiconductors just above their plasma frequency [21], so that such materials may also be used instead of the composite ENZ metamaterials in AC driven Coulomb blockade devices. For example, the dielectric permittivity of silicon carbide at 29 THz is $\varepsilon=0+i0.1$, leading to Coulomb blockade enhancement by approximately factor of 10. Such a material choice would also automatically satisfy the usual requirement that the metamaterial structural parameter must be much smaller than the typical device scale, so that the macroscopic electrodynamics description of the device remains valid.

Turning now to anisotropic metamaterial designs, one of the most interesting cases would be enhancement of Coulomb blockade in hyperbolic metamaterials [3]. Hyperbolic metamaterials are extremely anisotropic uniaxial materials, which behave like a metal in one direction and like a dielectric in the orthogonal direction. They are typically made of either metallic nanowire or nanolayer arrays. In the hyperbolic metamaterial scenario the effective Coulomb potential from Eq. (1) assumes the form

$$V(\vec{q},\omega) = \frac{4\pi e^2}{q_z^2 \varepsilon_2(\vec{q},\omega) + (q_x^2 + q_y^2)\varepsilon_1(\vec{q},\omega)} = \frac{4\pi e^2}{q^2(\varepsilon_2(\vec{q},\omega)\cos^2\alpha + \varepsilon_1(\vec{q},\omega)\sin^2\alpha)} \quad (8)$$

where $\varepsilon_{xx} = \varepsilon_{yy} = \varepsilon_1$ and $\varepsilon_{zz} = \varepsilon_2$ have opposite signs, and $\alpha$ is the angle between the direction of $q$ and the z-axis. As a result, the effective Coulomb interaction of two electrons may become repulsive and very strong along spatial directions where

$$q_z^2 \varepsilon_2(\vec{q},\omega) + (q_x^2 + q_y^2)\varepsilon_1(\vec{q},\omega) \approx 0 \quad (9)$$

while staying positive. On the other hand, just across the borderline described by Eq.(9) where the denominator of Eq.(8) becomes negative, a very strong attraction of electrons may occur. This opposite situation was shown theoretically and experimentally [5] to lead to a superconducting electronic state of the metamaterial, with the superconducting critical temperature Tc greatly increased compared to the Tc of the parent material. Thus, it is clear that superconductivity and the Coulomb blockade-induced charge ordering in hyperbolic metamaterials will compete with each other. Such a competition is indeed observed in many high Tc cuprates [22], which often behave as natural hyperbolic materials [23]. For example, suppression of Tc in hole-doped La-based cuprates is well known to occur near x = 1/8 doping, coinciding with the organization of charge into stripe-like patterns with four lattice constants periodicity [24].

Fig.4 illustrates one of the possible charge ordering scenarios in an artificial hyperbolic metamaterial made of periodic layers of metal separated by dielectric. In a layered hyperbolic metamaterial made of layers of metal (with thickness $d_m$) separated by layers of dielectric (with thickness $d_d$) the $\varepsilon_1$ and $\varepsilon_2$ components of the dielectric tensor are typically calculated using the Maxwell-Garnett approximation as follows:

$$\varepsilon_1 = \delta\varepsilon_m + (1-\delta)\varepsilon_d \tag{10}$$

$$\varepsilon_2 = \frac{\varepsilon_m \varepsilon_d}{(1-\delta)\varepsilon_m + \delta\varepsilon_d} \tag{11}$$

where $\delta = d_m/(d_m+d_d)$ is the volume fraction of metal, and $\varepsilon_m$ and $\varepsilon_d$ are the dielectric permittivities of the metal and dielectric, respectively [25]. In hyperbolic metamaterial superconductors the typical metal and dielectric layer are several nanometers thick [5], so that the average inter-carrier distance may be comparable with the metamaterial periodicity:

$$a = n^{-1/3} \propto (d_m + d_d) \tag{12}$$

Therefore as illustrated in Fig.4, a Mott-like transition to a charge-ordered state may potentially occur in a hyperbolic metamaterial in such a way that the maxima of the spatial charge distribution would be aligned along the positive ENZ direction defined by Eq.(9). It is clear that under conditions set by Eq.(12) the simple macroscopic electrodynamic result given by Eqs.(10,11) may need corrections. This conclusion is relevant not only for hyperbolic metamaterial superconductors, but for many semiconducting hyperbolic metamaterial geometries as well (see for example [26]). Experimental observation of charge ordering in a hyperbolic metamaterial geometry under the conditions set by Eq.(12) may also clarify the physics of charge ordering in natural high Tc cuprates.

As far as the hyperbolic metamaterial-based Coulomb blockade devices are concerned, the design of such devices may also take into account the described anisotropy of the electron-electron interaction. This may be done in a straightforward manner due to directionality of current in typical Coulomb blockade devices, as illustrated in Fig.5. The Coulomb blockade will be strongly enhanced if the direction of current is made to coincide with the positive ENZ direction set by Eq.(9). Using Eqs.(8-11) we have calculated expected Coulomb blockade enhancement factors in the THz range for a graphene/$Al_2O_3$ hyperbolic metamaterial described in [27]. Graphene-based plasmonic and metamaterial devices are known to perform well in a very broad 10-120 THz frequency range. The real and imaginary parts of the hyperbolic metamaterial permittivities $\varepsilon_1$ and $\varepsilon_2$ at 37 THz have been extracted from Fig.5c of [27]. The calculated enhancement factor as a function of angle is shown in Fig.5b. Note that at angles above 27 degrees the effective Coulomb interaction changes sign. Even larger enhancement factors of the order of 10 may be expected in the SiC example at 29 THz mentioned above, where the imaginary part of $\varepsilon$ may be as small as 0.1 [21]. Such a large Coulomb blockade enhancement will lead to ten times higher operating temperature range of the device.

In conclusion, we have analyzed several promising experimental geometries, which may considerably enhance performance of Coulomb blockade-based optoelectronic devices. We have demonstrated that positive ENZ conditions enhance Coulomb repulsion of charges, leading to considerable enhancement of the Coulomb blockade effect in ENZ and hyperbolic metamaterials. Similar to high Tc cuprates, in hyperbolic metamaterials superconductivity and the Coulomb blockade-induced charge ordering compete with each other. We also note that the enhancement of Coulomb blockade effect in positive ENZ metamaterials considered here, and the recently predicted levitation of electric dipoles above an ENZ metamaterial [28] have common physical origins. Both phenomena rely on the expulsion of the time-varying electric field from the metamaterial interior.

**Acknowledgement**

This work was supported in part by the DARPA Award No: W911NF-17-1-0348 "Metamaterial Superconductors".

**References**

[1] I. I. Smolyaninov, V. N. Smolyaninova, *Phys. Rev. B* **93**, 184510 (2016).

[2] N. Engheta, *Science* **340**, 286 (2013).

[3] Z. Jakob, L.V. Alekseyev, E. Narimanov, *Optics Express* **14,** 8247 (2006).

[4] V.N. Smolyaninova, K. Zander, T. Gresock, C. Jensen, J.C. Prestigiacomo, M.S. Osofsky, I. I. Smolyaninov, *Scientific Reports* **5**, 15777 (2015).

[5] V.N. Smolyaninova, C. Jensen, W. Zimmerman, J.C. Prestigiacomo, M.S. Osofsky, H. Kim, N. Bassim, Z. Xing, M. M. Qazilbash, I.I. Smolyaninov, *Scientific Reports* **6**, 34140 (2016).

[6] P. Delsing, K. K. Likharev, L. S. Kuzmin, T. Claeson, *Phys. Rev. Lett*. **63**, 1180 (1989).

[7] M. H. Devoret, D. Esteve, H. Grabert, G.-L. Ingold, H. Pothier, C. Urbina. *Phys. Rev. Lett.* **64**, 1824 (1990).

[8] S. M. Girvin, L. I. Glazman, M. Jonson, D. R. Penn and M. D. Stiles, *Phys. Rev. Lett.* **64**, 3183 (1990).

[9] A. Altland, A. De Martino, R. Egger, B. Narozhny, *Phys. Rev. B* **82**, 115323 (2010).

[10] I. Sa, E. V. Sukhorukov, *EPL* **91**, 67008 (2010).

[11] H. Grabert, *Phys. Rev. B* **92**, 245433 (2015).

[12] K. Shibata, A. Umeno, K. M. Cha and K. Hirakawa, *Phys. Rev. Lett*. **109**, 077401 (2012).

[13] G. Gasse, L. Spietz, C. Lupien and B. Reulet, *Phys. Rev. B* **88**, 241402 (2013).

[14] S. J. Shin, J. J. Lee, H. J. Kang, J. B. Choi, S.-R. Yang, Y. Takahashi, D. G. Hasko, *Nano Letters* **11**, 1591-1597 (2011).

[15] V.N. Smolyaninova, B. Yost, K. Zander, M. S. Osofsky, H. Kim, S. Saha, R. L. Greene, I. I. Smolyaninov, *Scientific Reports* **4**, 7321 (2014).

[16] T.C. Choy, *Effective Medium Theory* (Clarendon Press, Oxford, 1999).

[17] A. Alù, N. Engheta, *Phys. Rev. E* **72**, 016623 (2005).

[18] C. Kittel, *Introduction to Solid State Physics* (Wiley, New York, 2004).

[19] R. E. Lindquist, A. W. Ewald, *Physical Review* **135**, A191-A194 (1964).

[20] M. P. McNeal, S.-J. Jang, R. E. Newnham, *J. of Appl. Phys.* **83**, 3288-3297 (1998).

[21] I. Liberal, N. Engheta, *Nature Photonics* **11**, 149–158 (2017).

[22] E. H. da Silva Neto, P. Aynajian1, A. Frano, R. Comin, E. Schierle, E. Weschke, A. Gyenis, J. Wen, J. Schneeloch, Z. Xu, S. Ono, G. Gu, M. Le Tacon, A. Yazdani, *Science* **343**, 393-396 (2014).

[23] I. I. Smolyaninov, *Journal of Physics: Condensed Matter* **26**, 305701 (2014).

[24] J. M. Tranquada, B. J. Sternlieb, J. D. Axe, Y. Nakamura, S. Uchida, *Nature* **375**, 561 (1995).

[25] R. Wangberg, J. Elser, E. E. Narimanov, V. A. Podolskiy, , *J. Opt. Soc. Am. B* **23**, 498 (2006).

[26] S. Campione, S. Liu, T. S. Luk, M. B. Sinclair, *J. Opt. Soc. Am. B* **32**, 1809-1815 (2015).

[27] Y.-C. Chang, C.-H. Liu, C.-H. Liu, S. Zhang, S. R. Marder, E. E. Narimanov, Z. Zhong, T. B. Norris, *Nature Communications* **7**, 10568 (2016).

[28] F. J. Rodríguez-Fortuño, A. Vakil, N. Engheta, *Phys. Rev. Lett*. **112**, 033902 (2014).

**Figure Captions**

**Figure 1.** Schematic geometry of conventional (a) and positive ENZ (b) Coulomb blockade device. When the island between the source and the drain is surrounded by the positive ENZ metamaterial, its self-capacitance is greatly reduced. Note that in the AC driven Coulomb blockade device the source and the drain may be integrated into an antenna structure, such as a bowtie antenna, see for example ref. [12].

**Figure 2.** (a) SEM image of a metal-dielectric metamaterial based on random mixing of tin and $BaTiO_3$ nanoparticles. Individual nanoparticles are clearly visible in the image. (b) Schematic view of core-shell nanoparticle geometry based on dielectric core and metal shell. Such core-shell nanoparticles may be used to fabricate broadband positive ENZ metamaterials.

**Figure 3.** (a) Numerical calculations of the real part of the dielectric constant of the $Sn/BaTiO_3$ mixture as a function of volume fraction of $BaTiO_3$. (b) The plots of the real part of $\varepsilon$ for pure tin and for the positive ENZ tin-$BaTiO_3$ nanocomposite metamaterial. The volume fraction of $BaTiO_3$ in the metamaterial is 40%.

**Figure 4**. One of the possible charge ordering scenarios in an artificial hyperbolic metamaterial made of periodic layers of metal separated by dielectric. The color scheme is chosen to illustrate the spatial distribution of charge inside the individual metal layers. The charge-ordered state is favored when the peaks of charge density are aligned along the positive ENZ direction.

**Figure 5**. (a) Schematic geometry of a Coulomb blockade device using a hyperbolic metamaterial. The positive ENZ direction is indicated by red arrow. The Coulomb blockade will be strongly enhanced if the direction of current is made to coincide with the positive ENZ direction set by Eq.(9). (b) Calculated Coulomb blockade

enhancement factor as a function of angle at 37 THz for a graphene/$Al_2O_3$ hyperbolic metamaterial described in [27]. Above 27 degrees the effective Coulomb interaction changes sign.

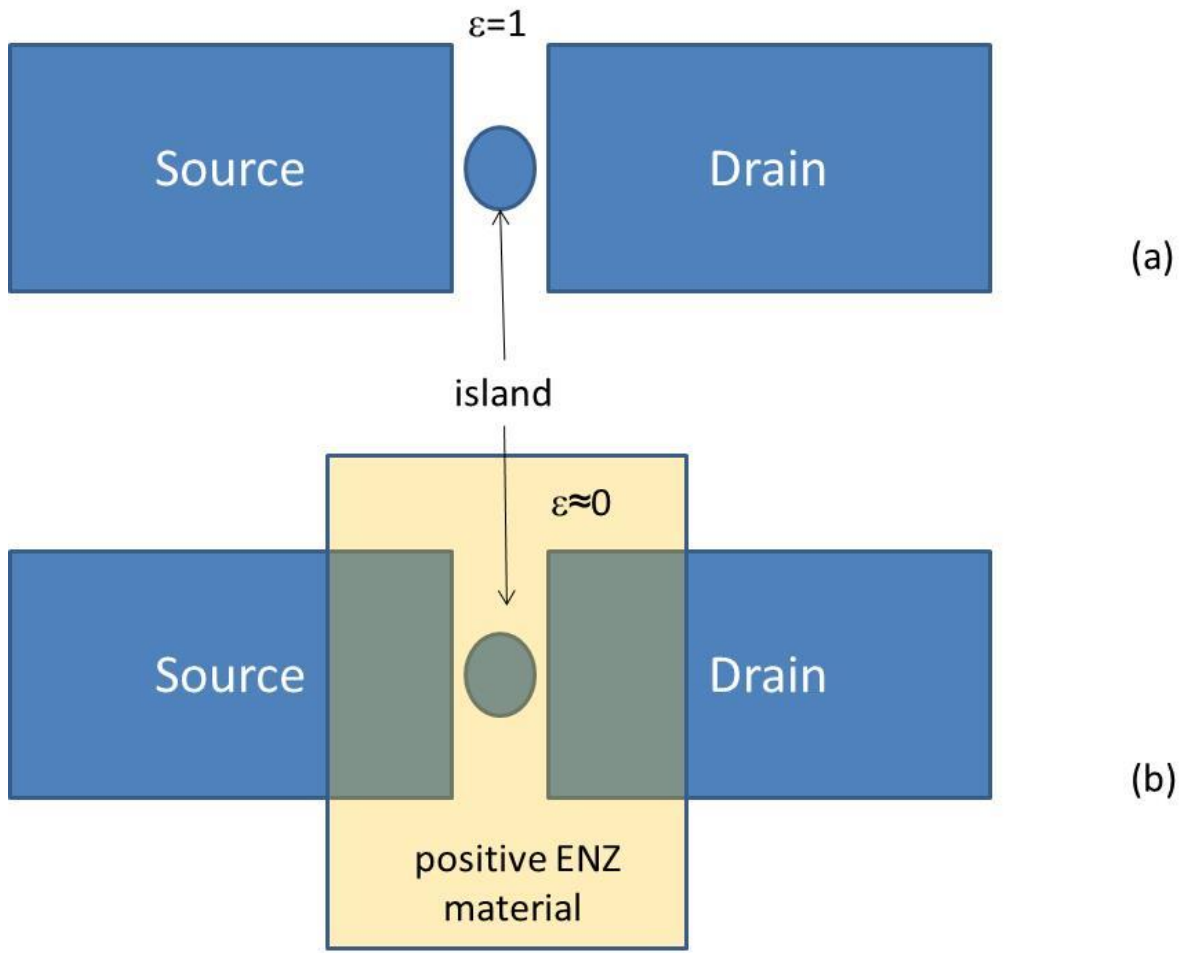

Fig.1

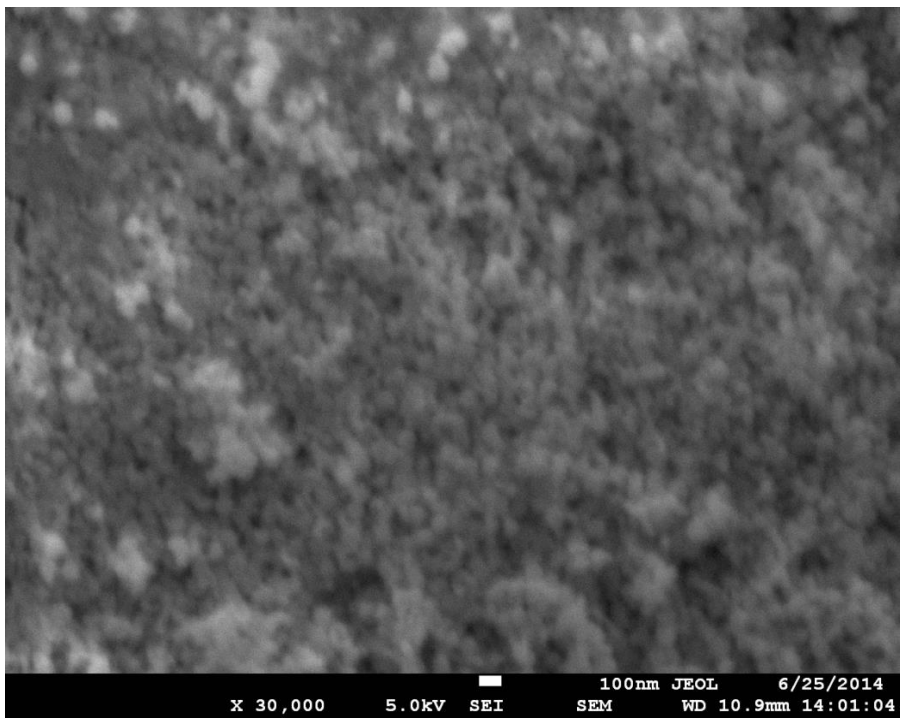

(a)

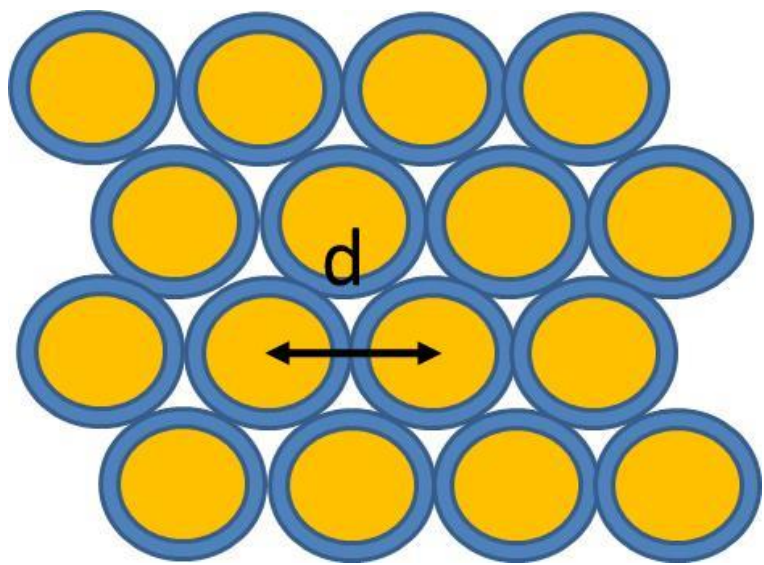

(b)

Fig.2

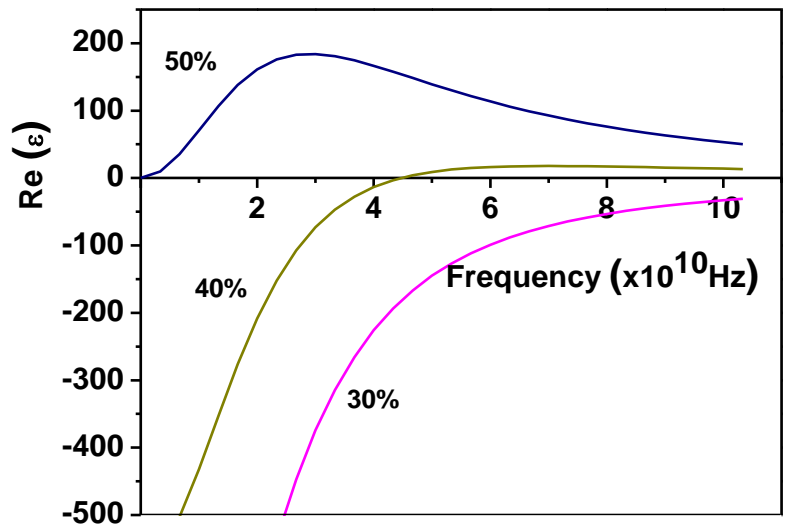

(a)

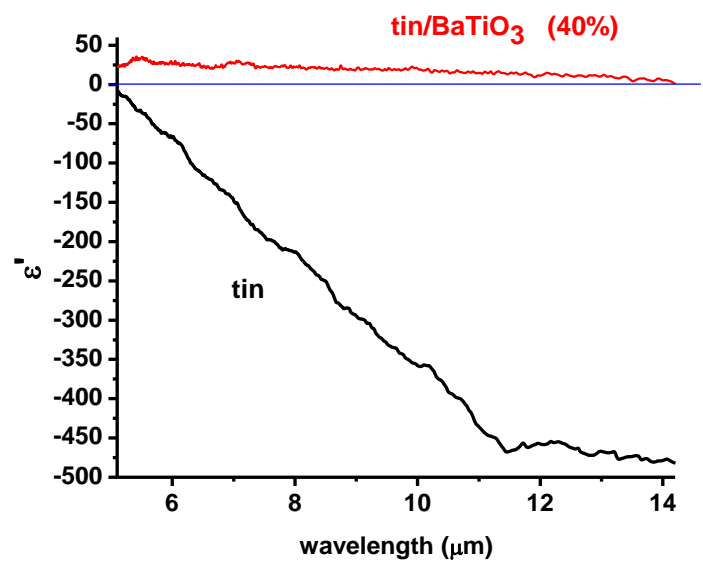

(b)

Fig. 3

positive $\varepsilon \approx 0$ directions

$d_m$ metal layers

$d_d$

dielectric layers

$a \sim n^{-1/3}$

volume fraction of metal: $\delta = d_m/(d_m+d_d)$

Fig. 4

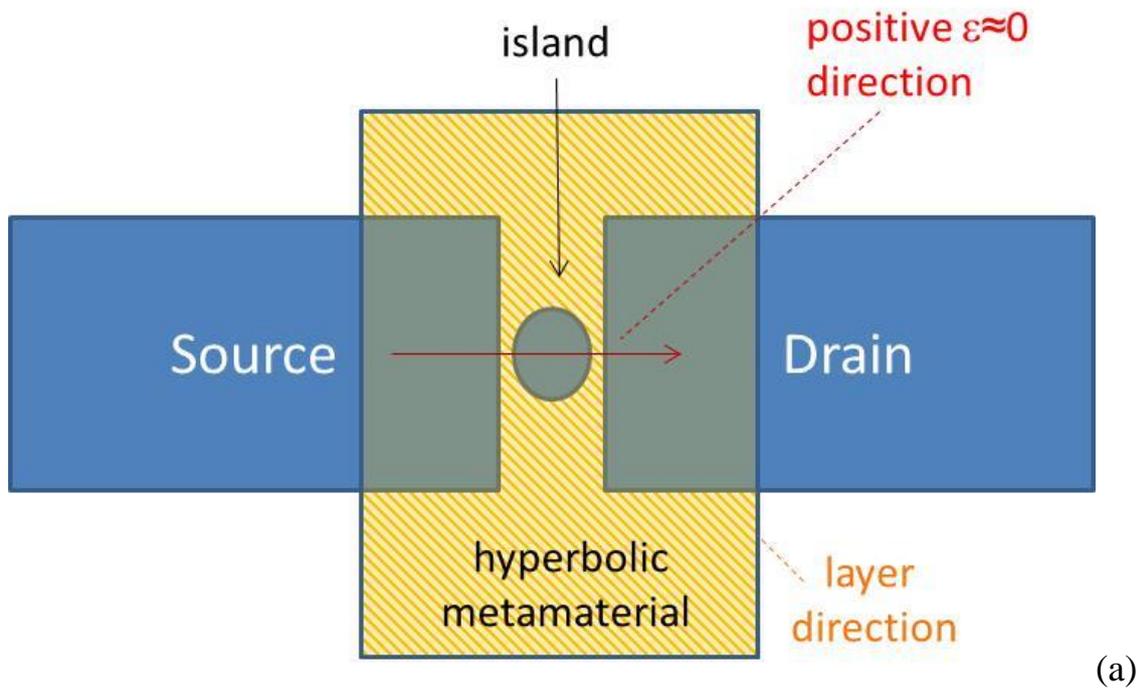

(a)

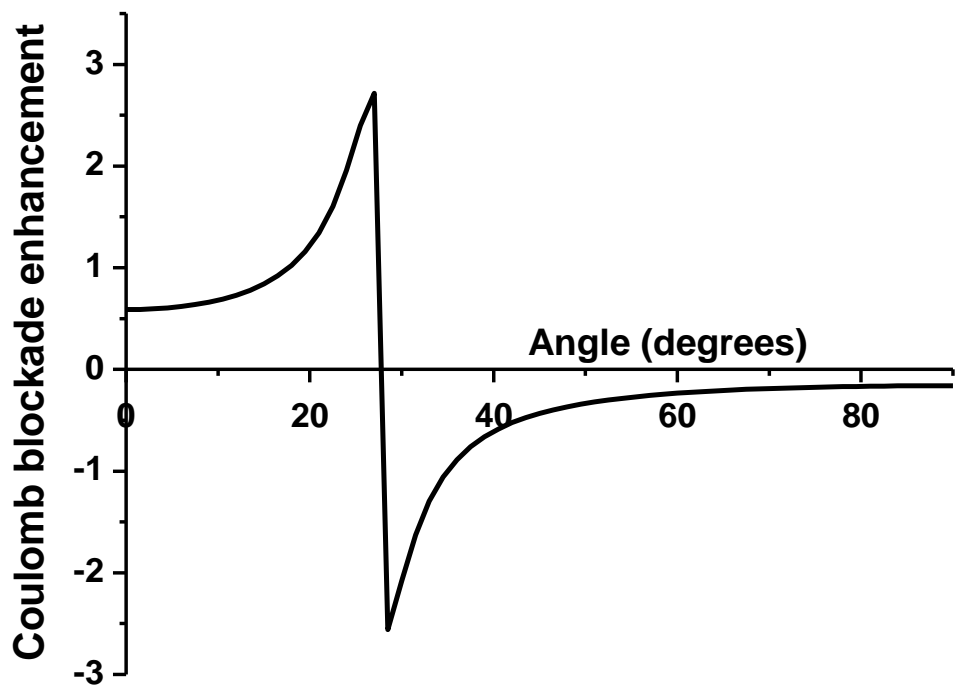

(b)

Fig. 5